\documentclass[a4paper]{article}

\usepackage{INTERSPEECH2022}
\usepackage{amsmath}
\usepackage{amssymb}
\usepackage{mathtools}
\usepackage{amsfonts}       
\usepackage{bbm}
\usepackage{url}            
\usepackage{microtype}      
\usepackage{multirow}
\usepackage{float}
\usepackage{caption}
\usepackage{subcaption}
\usepackage{hyperref}
\usepackage{graphicx}
\usepackage{tikz}
\usetikzlibrary{automata,positioning}

\title{Low-resource Low-footprint Wake-word Detection using Knowledge Distillation}
\name{Arindam Ghosh\sthanks{\ \ Equal contribution.}, Mark Fuhs\footnotemark[1], Deblin Bagchi, Bahman Farahani, Monika Woszczyna}

\address{3M Health Information Systems}
\email{\{aghosh4, mark.fuhs, dbagchi, bmashhadifarahani, mwoszczyna\}@mmm.com}

\begin{document}

\maketitle
\begin{abstract}
As virtual assistants have become more diverse and specialized, so has the demand for application or brand-specific wake words. However, the wake-word-specific datasets typically used to train wake-word detectors are costly to create. In this paper, we explore two techniques to leverage acoustic modeling data for large-vocabulary speech recognition to improve a purpose-built wake-word detector: transfer learning and knowledge distillation. We also explore how these techniques interact with time-synchronous training targets to improve detection latency. Experiments are presented on the open-source “Hey Snips” dataset and a more challenging in-house far-field dataset. Using phone-synchronous targets and knowledge distillation from a large acoustic model, we are able to improve accuracy across dataset sizes for both datasets while reducing latency.
\end{abstract}

\section{Introduction}
\label{sec:intro}

Speech interfaces for virtual assistants typically use a wake word to initiate interaction with the assistant. In recent years, virtual assistants have become more popular but also more diverse, including specialized applications in such areas as automotive and healthcare. Typically, it is important to use a custom or brand-specific wake word (e.g., the name of an automobile’s manufacturer) to tie the assistant to the product into which it is embedded, even if the underlying virtual assistant technology is licensed from another company. 

To minimize bandwidth and response latency, wake-word detectors typically run on the embedded or mobile device hardware proximal to the user, constraining the  computing footprint of the model. To maximize performance, such models are typically trained on purpose-built wake-word-specific datasets; however, these data resources are costly. We explore two approaches to improve the accuracy of wake-word detectors using datasets intended for training large-vocabulary speech recognition acoustic models: transfer learning and knowledge distillation.

In parameter- or model-based transfer learning, a network is first trained on a related task, then retrained on or reused for the main task (see \cite{zhuang2021} for review). The approach has been applied in the wake-word setting using Automatic Speech Recognition (ASR) acoustic modeling as the pretraining task \cite{chen2014,sun2017,gao2020}. In particular, \cite{chen2014} explores keyword spotting accuracy with and without transfer learning, though they do not examine the impact of transfer learning at different keyword-specific dataset sizes.

Knowledge distillation \cite{hinton2015distilling} is a popular approach for model compression that has been used in speech recognition \cite{lu2017knowledge,bagchi2019learning,takashima2018investigation,kurata2018improved}. In the keyword spotting literature, \cite{tucker2016model} explores the use of an ensemble of wake-word model teachers for model compression using a large corpus of positive utterances (est. $>400$K). Similarly, \cite{dighe2021} uses a combined loss of knowledge distillation from teacher's LatticeGNN embeddings and the main wake-word classification task to train a small student network, again focusing on the large data setting, with more than 1 million positive utterances. Finally, \cite{park2021_interspeech} uses  teacher-student training to iteratively improve a student model using large amounts of labeled (2.5M utterances) and unlabeled data (10M utterances).


In contrast, we focus on the low-resource setting, where we explore how to improve the accuracy and latency of a strong baseline system \cite{wang2020} when wake-word data is limited. In Section \ref{sec:overview}, we describe the datasets and baseline system, as well as the use of phone-aligned training to improve latency. Section \ref{sec:traindesc} details the multi-stage training approaches, and, in Section \ref{sec:exp}, we provide experimental results to systematically compare wake-word-only training with transfer learning and knowledge distillation on different dataset sizes from two wake-word datasets, one with isolated wake words (Snips) and one with wake-word-prefixed virtual assistant requests (Fluency).


\vspace{-0.1cm}
\section{System Overview}
\label{sec:overview}

\subsection{Datasets}
\label{sec:datasets}
Wake-word experiments are carried out on two datasets: (1) the publicly available Snips dataset \cite{snips2019}, consisting of the wake word ``Hey Snips" spoken alone; and (2) an in-house Fluency dataset consisting of the wake words ``Hey Fluency" and ``Okay Fluency" followed by a request to the digital assistant, e.g., ``Hey Fluency who is the next patient?'' Table \ref{tab:dataset} presents a summary of the datasets.

For the Fluency dataset, positive training examples were recordings from near-field microphones, while a limited set of far-field recordings are used for the test set. Non-wake-word data is taken from a large in-house corpus of far-field conversational speech. The far-field audio quality and lack of isolation of the wake word make the Fluency test set more challenging.

\begin{table}[hb]
    \vspace{-0.2cm}
    \caption{Datasets. The positive examples are given in number of utterances whereas the negative examples are given in hours.}
    \vspace{-0.2cm}
    \label{tab:dataset}
    \centering
    \resizebox{\linewidth}{!}{
      \begin{tabular}{ c|c c|c c }
        \hline
        \multicolumn{1}{c|}{\textbf{Dataset}} & \multicolumn{2}{c|}{\textbf{Train}}  & \multicolumn{2}{c}{\textbf{Eval}}\\
                    &Positive &Negative   &Positive &Negative\\
        \hline
        Snips   &5,799 utt  &50.64h         &2,529 utt  &23.19h\\
        \hline
        Fluency     &7,169 utt  &153.97h         &3,840 utt &434.37h\\
    
        \hline
      \end{tabular}
  }
\end{table}
\vspace{-0.15cm}

To simulate low resource settings, we take the first $n$ examples to create training subsets containing 100, 500, 1000, and 2000 positive utterances. For Snips, the subsets contain utterances from 22, 106, 203, and 404 speakers, respectively, and the full training set of 5799 utterances comprises 1163 speakers. For Fluency, subsets 100, 500, 1000, and 2000 contain utterances from 19, 78, 86, and 105 speakers respectively. The full training set of 7169 utterances comprises 121 speakers.
    

In contrast to the small wake word datasets, a separate in-house 2800h near-field dataset is used to train the acoustic models that are used for transfer learning and as the teacher network for knowledge distillation (Section \ref{sec:traindesc}).

\subsection{Lattice-free Maximum Mutual Information}
\vspace{-0.1cm}
In this work, for training or fine-tuning we use either the regular or the alignment-free variant of the lattice-free maximum mutual information (LF-MMI) objective \cite{povey2016_lfmmi}, which is given by
\begin{equation}
    \mathcal{F}_{\mathrm{LF-MMI}} = \sum_{n=1}^{N}\log \mathbf{P}(L_n | O_n) = \sum_{n=1}^{N}\log \frac{\mathbf{P}(O_n | L_n)\mathbf{P}(L_n)}{\sum_{L} \mathbf{P}(O_n|L)\mathbf{P}(L)}
    \label{eq:lfmmi_loss}
\end{equation}
where $O_n$ is the input audio and $L_n$ and $L$ are the true and competing hypothesis sequences respectively. 

For ASR tasks, the numerator graph in alignment-free LF-MMI is constructed as an unexpanded graph (with self loops) using the training transcripts \cite{hadian18_interspeech}. Like Connectionist Temporal Classification (CTC) loss, this constrains the sequence of senone targets without explicit time information. In contrast, for regular LF-MMI, a prior acoustic model is force-aligned with the data to yield time constrains that are then represented in the numerator lattice using an acyclic and expanded graph (no self-loops) with each path in the lattice having one state per frame. This constrains the training targets with explicit time information.

The denominator graph for regular LF-MMI is constructed using a phone language model (LM) trained on the phone alignments of the training data. For alignment-free LF-MMI, since there's no alignment information available for the training data, the phone LM is estimated from the training transcripts by including random pronunciations for the words that have multiple pronunciations and inserting silence at the beginning, end, and between the words with some probabilities. 

\subsection{Baseline System}
\label{sec:baselinesys}
\vspace{-0.1cm}
As the baseline system, we use a state-of-the-art TDNN-F/HMM system trained with alignment-free LF-MMI \cite{wang2020}. It uses left-to-right 4-state HMM “chain” topologies to model the wake-word and general speech, and a 1-state HMM topology to model silence. The network has two outputs per state: one for the likelihood of the transition into the state, the other for the likelihood of the state’s self-loop. Similar to ASR training, the numerator graph is an unexpanded FST constructed from the transcript: either “WakeWord” (Snips dataset) or “WakeWord Speech" (Fluency dataset) for positive utterances, and “Speech” for negative utterances. The denominator graph, however, is a manually specified topology that comprises paths with and without the wake-word. The baseline system is implemented in the Kaldi toolkit \cite{kaldi2011}, which we also use for our experiments.


For data augmentation, we apply the techniques used in \cite{wang2020}, including simulated reverberation \cite{ko2017}, speed perturbation \cite{ko15_interspeech} and noise, music, and background speech from the MUSAN corpus \cite{musan2015}.
For Snips dataset, we use the mentioned augmentations for both positive and negative examples. For Fluency dataset, however, we augment only the positive examples, as the negative examples are already from far-field recordings.

\subsection{Phone-aligned Training}
\vspace{-0.1cm}
As an alternative to alignment-free training, we explore phone-aligned numerator lattices. While allowing the network to settle on its own alignment to the data is likely optimal for accuracy in large-data contexts, we hypothesized that the additional time information would improve accuracy especially when the data is limited or more challenging. Moreover, while alignment-free training focuses solely on accuracy, constraining the model’s output in time would allow for reduction in latency.

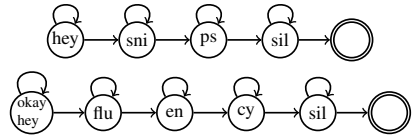
\begin{figure}[t]
    \centering
    \resizebox{0.8\linewidth}{!}{
    \begin{subfigure}[]{\linewidth}
        \centering
        \begin{tikzpicture}[node distance={12mm}, thick, main/.style = {draw, circle, minimum size=6mm, inner sep=0}]
            \node[main] (0) {hey};
            \node[main] (1) [right of=0] {sni};
            \node[main] (2) [right of=1] {ps};
            \node[main] (3) [right of=2] {sil};
            \node[main] (4) [right of=3] {};
            \node[main, minimum size=7mm] (5) [right of=3] {};
            \draw[->] (0) -- (1); 
            \draw[->] (1) -- (2);
            \draw[->] (2) -- (3);
            \draw[->] (3) -- (5);
            \draw[->] (0) to [out=120,in=60,looseness=4] (0);
            \draw[->] (1) to [out=120,in=60,looseness=4] (1);
            \draw[->] (2) to [out=120,in=60,looseness=4] (2);
            \draw[->] (3) to [out=120,in=60,looseness=4] (3);
        \end{tikzpicture}
        \label{fig:hey_snips_numerator_G.fst}
    \end{subfigure}
    }
    
    \resizebox{0.8\linewidth}{!}{
    \begin{subfigure}[]{\linewidth}
        \centering
        \begin{tikzpicture}[node distance={12mm}, thick, main/.style = {draw, circle, minimum size=6mm, inner sep=0}]
            \node[main, text width=6mm, scale=0.8] (0) {okay hey};
            \node[main] (1) [right of=0] {flu};
            \node[main] (2) [right of=1] {en};
            \node[main] (3) [right of=2] {cy};
            \node[main] (4) [right of=3] {sil};
            \node[main] (5) [right of=4] {};
            \node[main, minimum size=7mm] (6) [right of=4] {};
            \draw[->] (0) -- (1); 
            \draw[->] (1) -- (2);
            \draw[->] (2) -- (3);
            \draw[->] (3) -- (4);
            \draw[->] (4) -- (6);
            \draw[->] (0) to [out=120,in=60,looseness=4] (0);
            \draw[->] (1) to [out=120,in=60,looseness=4] (1);
            \draw[->] (2) to [out=120,in=60,looseness=4] (2);
            \draw[->] (3) to [out=120,in=60,looseness=4] (3);
            \draw[->] (4) to [out=120,in=60,looseness=4] (4);
        \end{tikzpicture}
        \label{fig:hey_fluency_hmm}
    \end{subfigure}
    }
    
    \caption{Wake-word HMM topologies for``Hey Snips" (top) and ``Hey/Okay Fluency" (bottom).}
    \label{fig:hmm_topology}
    \vspace{-0.6cm}
\end{figure}

We introduce the automatically-inferred time constraints in these lattices using the forced alignment of the wake-words from an in-house ASR model to infer phone-level frame labels. Instead of assigning an HMM state to each phone, we cover a group of phones with a single HMM state. This reduces the number of outputs in the last layer, and thus the size, of the neural network. Fig. \ref{fig:hmm_topology} shows the HMM states and the partition of phones across them for the wake-words in both datasets. To account for the Hey/Okay ambiguity, we assign the first state of the wake-word HMM to cover the time span of either word. To model the pause that typically occurs after the wake word is spoken, we assign the silence that immediately follows the end of the wake-word (denoted by ``sil") to the last state of the wake-word HMM. This last state is trained to model the first 10 frames (100 ms) of post-wake-word silence, while any other silence/non-speech that follows is covered by the silence HMM. In early experiments, we found this post-wake-word ``sil" state to reduce false positives.

As in the baseline system, general speech is modeled using the same HMM topology as that of the corresponding wake-word, and silence is modeled by the 1-state HMM mentioned before. To achieve phonetic alignment with the audio, time constraints must therefore be imposed throughout the numerator lattice. In early experiments, we found good performance simply by spreading the HMM states of the general speech equally over the duration of the general speech region. 

\begin{figure}[ht]
    \vspace{-0.1cm}
            
            
            
    
    \resizebox{0.8\linewidth}{!}{
    \begin{subfigure}[]{0.5\linewidth}
        \begin{tikzpicture}[node distance={12mm}, thick, main/.style = {draw, circle}]
            \node[main] (0) {0};
            \node[main] (1) [right of=0] {1};
            \node[main] (2) [right of=1] {2};
            \node[main] (3) [right of=2] {3};
            \node[main] (4) [below of=2] {4};
            \node[main] (5) [right of=4] {5};
            
            \node[main, minimum size=6.75mm] (6) [right of=2] {}; 
            \node[main, minimum size=6.75mm] (7) [right of=4] {}; 
            
            \draw[->] (0) -- node[midway, above, sloped, pos=0.5] {SIL} (1);
            \draw[->] (1) -- node[midway, above, sloped, pos=0.5] {WW} (2);
            \draw[->] (2) -- node[midway, above, sloped, pos=0.5] {SIL} (6);
            \draw[->] (1) -- node[midway, above, sloped, pos=0.5] {Speech} (4);
            \draw[->] (4) -- node[midway, above, sloped, pos=0.5] {SIL} (7);
            \draw[->] (0) to [out=270,in=225,looseness=0.8] node[midway, above, sloped, pos=0.5] {SIL} (7);
        \end{tikzpicture}
    \label{fig:hey_snips_phone_lm}
    \end{subfigure}
    
    \begin{subfigure}[]{0.49\linewidth}
        \vspace{-0.7cm}
        \hspace*{2.5cm} \begin{tikzpicture}[node distance={12mm}, thick, main/.style = {draw, circle}]
            \node[main] (0) {0};
            \node[main] (1) [right of=0] {1};
            \node[main] (2) [right of=1] {2};
            \node[main] (4) [below of=2] {4};
            
            \node[main, minimum size=6.75mm] (6) {}; 
            
            \draw[->] (6) to [out=300,in=190,looseness=0.8] node[midway, above, pos=0.5] {SIL} (4);
            \draw[->] (0) -- node[midway, above, sloped, pos=0.5] {SIL} (1);
            \draw[->] (1) -- node[midway, above, sloped, pos=0.5] {WW} (2);
            \draw[->] (4) to [out=220,in=270,looseness=0.8] node[midway, below, sloped, pos=0.5] {SIL} (6);
            \draw[->] (1) -- node[midway, above, sloped, pos=0.5] {Speech} (4);
            \draw[->] (2) to [out=90,in=60,looseness=0.8] node[midway, above, sloped, pos=0.5] {SIL} (6);
            
        \end{tikzpicture}
        \label{fig:hey_snips_decoding_G.fst}
    \end{subfigure}
    }
            
            
            
    \vspace{-0.7cm}
    \caption{Denominator graph (left) and decoding graph (right) for ``Hey Snips".}
    \label{fig:hey_snips_graphs}
    \vspace{-0.2cm}
\end{figure}
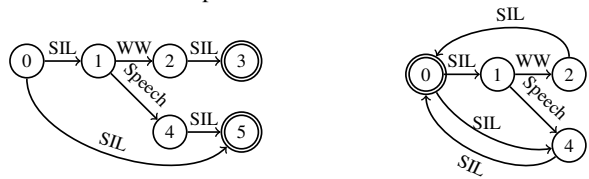

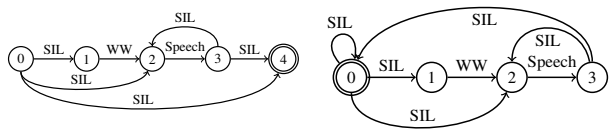
\begin{figure}[H]
    \vspace{-0.8cm}
    \centering
    
    \begin{subfigure}[]{0.49\linewidth}
        \centering
        \resizebox{\linewidth}{!}{
        \begin{tikzpicture}[node distance={15mm}, thick, main/.style = {draw, circle}]
            \node[main] (0) {0};
            \node[main] (1) [right of=0] {1};
            \node[main] (2) [right of=1] {2};
            \node[main] (3) [right of=2] {3};
            \node[main] (4) [right of=3] {4};
        
            \node[main, minimum size=6.75mm] (6) [right of=3] {}; 
            
            \draw[->] (0) -- node[midway, above, sloped, pos=0.5] {SIL} (1);
            \draw[->] (1) -- node[midway, above, sloped, pos=0.5] {WW} (2);
            \draw[->] (2) -- node[midway, above, sloped, pos=0.5] {Speech} (3);
            \draw[->] (3) -- node[midway, above, sloped, pos=0.5] {SIL} (6);
            \draw[->] (3) to [out=90,in=90,looseness=1] node[midway, above, pos=0.5] {SIL} (2);
            \draw[->] (0) to [out=270,in=250,looseness=0.5] node[midway, above, sloped, pos=0.5] {SIL} (2);
            \draw[->] (0) to [out=270,in=250,looseness=0.5] node[midway, above, sloped, pos=0.5] {SIL} (6);
        \end{tikzpicture}
        }
    \label{fig:hey_fluency_phone_lm}
    \end{subfigure}
    \begin{subfigure}[]{0.5\linewidth}
        \centering
        \resizebox{\linewidth}{!}{
        \begin{tikzpicture}[node distance={15mm}, thick, main/.style = {draw, circle}]
            \node[main] (0) {0};
            \node[main] (1) [right of=0] {1};
            \node[main] (2) [right of=1] {2};
            \node[main] (3) [right of=2] {3};
        
            \node[main, minimum size=6.75mm] (6) {}; 
            
            \draw[->] (0) -- node[midway, above, sloped, pos=0.5] {SIL} (1);
            \draw[->] (1) -- node[midway, above, sloped, pos=0.5] {WW} (2);
            \draw[->] (2) -- node[midway, above, sloped, pos=0.5] {Speech} (3);
            \draw[->] (3) to [out=100,in=90,looseness=1.5] node[midway, below, pos=0.5] {SIL} (2);
            \draw[->] (3) to [out=90,in=70,looseness=0.8] node[midway, below, pos=0.5] {SIL} (6);
            \draw[->] (0) to [out=270,in=250,looseness=0.8] node[midway, above, sloped, pos=0.5] {SIL} (2);
            \draw[->] (6) to [out=130,in=80,looseness=6.5] node[midway, above, pos=0.5] {SIL} (6);
        \end{tikzpicture}
        }
    \label{fig:hey_fluency_decoding_G.fst}
    \end{subfigure}
    \vspace{-0.7cm}
    \caption{Denominator graph (left) and decoding graph (right) for ``Hey/Okay Fluency Speech".}
    \label{fig:fluency_graphs}
    \vspace{-0.2cm}
\end{figure}

The “grammar” for the denominator graph are shown in Fig. \ref{fig:hey_snips_graphs} (Snips) and Fig. \ref{fig:fluency_graphs} (Fluency), which includes the wake-word and non-wake-word paths.

\subsection{Decoding}
\vspace{-0.1cm}
The decoding graph for Snips dataset is given in Fig. \ref{fig:hey_snips_graphs} and for Fluency in Fig. \ref{fig:fluency_graphs}. These are similar to the denominator graphs but assume a continuous audio stream comprising general speech and potentially many instances of wake words.

\vspace{-0.1cm}
\subsection{Neural Network Architecture}
\vspace{-0.1cm}
Since the focus of the current work is to explore training strategies to compensate for small wake-word training sets, experiments use a single general TDNN-F network architecture, similar to \cite{wang2020}, though we explore variations (layer size, number of layers, time strides, etc.) to the improve performance of each condition. A TDNN-F network is a TDNN network with its weight matrices in each layer factorized into the product of two low-rank matrices (the first matrix is semi-orthogonal) to reduce the number of parameters \cite{povey18_interspeech}. The skip connections are similar to those found in ResNets \cite{He_2016_CVPR} where each TDNN-F layer's output is added to the output from its previous layer (scaled by 0.66) before feeding into the next layer. To reduce latency, the time offsets of most of the TDNN layers are configured to be historical (looking at the prior layer's input at the current time step and prior time steps only) in order to limit the network's overall dependence on future frames to no more than 10. In order to reduce computation, the output frames are evaluated every 3 frames for computing LF-MMI loss during training and also during inference. In our low footprint settings, we keep the number of parameters to less than 400k for all our models. 

Input features are 64-dim log Mel filter banks extracted from the audio using a 23ms window with a 10ms frame shift. From the HMM topologies described earlier, the number of targets is 18 for the Snips HMM and 22 for the Fluency HMM.

\vspace{-0.1cm}
\section{Multi-stage Training}
\label{sec:traindesc}

\subsection{Transfer Learning}
\vspace{-0.1cm}
In this approach, we explore how pre-trained acoustic models for speech recognition can be applied to recognize a single wake word. While an acoustic model suitable for speech recognition could be used as a highly accurate wake-word detector, such models are far too large for the low-footprint applications of wake-word detectors. We therefore trained small acoustic models with the same general architecture: 1 TDNN layer (+-2 frames), 5 TDNN-F layers, 1 RELU layer, 9 TDNN-F layers, and the prefinal (1280-dim to 256-dim bottleneck) and final layers (4400 senones) typical of Kaldi chain training recipes. To reduce latency, networks looked ahead in time 10 frames, with most TDNN-F layers only looking back in time.

The first several layers of the acoustic model were transferred to the wake-word detector network, atop of which three additional TDNN-F layers (two with time strides -32, -16, 0, then -4, -2, 0) and the softmax layer were added with randomly initialized weights. Early experiments showed that using the first six layers of a network with 128-dim outer / 64-dim bottleneck TDNN-F layers outperformed using the first 16 layers of a network with 128-dim outer / 40-dim bottleneck TDNN-F layers; these alternatives had a similar number of parameters. Results are therefore reported on the 6-layer transfer. The weights of the transferred layers were fixed during fine-tuning as we found the model to lose performance when they were updated. 

\vspace{-0.1cm}
\subsection{Knowledge Distillation}
\begin{figure}[t]
    \centering
    \resizebox{0.6\linewidth}{!}{
    \begin{tikzpicture}
        \node[draw, rectangle, thick, minimum width = 2cm, minimum height = 4cm, text width=1cm] (r0) at (-1,0) {Teacher (17M)};
        \node[draw, rectangle, minimum width = 2cm, minimum height = 0.25cm] (r5) at (-1,1.78) {Hidden ($\mathbf{z}$)};
        
        \node[draw, rectangle, thick, minimum width = 2cm, minimum height = 1.5cm, text width=1cm] (r1) at (3,-1.25) {Student Lower (313K)};
        \node[draw, rectangle, minimum width=2cm, minimum height=0.25cm] (r6) at (3,-0.26) {Hidden ($\mathbf{\hat{z}}$)};
        \draw [-stealth, thick](-0,1.8) -- node[midway, below, sloped, pos=0.5, thick] {MSE Loss} (2,-0.25);
        
        \node[draw, rectangle, thick, minimum width = 2cm, minimum height = 0.5cm, text width=1.8cm] (r2) at (3,0.65) {Student\\Upper (55K)};
        \node[rectangle, thick, minimum width = 1.5cm, minimum height = 0.25cm, text width=2cm] (r7) at (3,1.75) {Wakeword\\LF-MMI Loss};
        \draw [-stealth, thick](3,1.4) -- (3,1.1);
        
        \node[draw, rectangle, thick, minimum width = 4.25cm, minimum height = 0.25cm] (r3) at (1,-2.75) {Acoustic Features ($\mathbf{x}$)};
        \draw [-latex, thick](1,-2.5) -- (-1,-2);
        \draw [-latex, thick](1,-2.5) -- (3,-2);
    \end{tikzpicture}
    }
    \vspace{-0.1cm}
    \caption{Teacher-student training setup. The number of model parameters is shown in parentheses.}
    \label{fig:teacher_student_setup}
    \vspace{-0.3cm}
\end{figure}
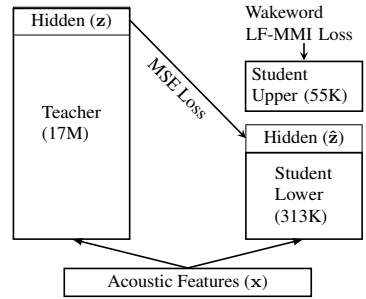

\vspace{-0.1cm}
In our knowledge distillation setup, as shown in Fig. \ref{fig:teacher_student_setup}, for an audio sample $\mathbf{x}$ (from the wake-word dataset), we use the teacher (a large ASR TDNN-F acoustic model) to generate hidden layer representations $\mathbf{z}$ from its penultimate bottleneck layer (128-dim), sized to match the corresponding student network. The output of the student network’s lower layers $\mathbf{\hat{z}}$ is regressed to the teacher representation via mean squared error (MSE) loss $\mathcal{L}_{\mathrm{MSE}} = \frac{1}{N} \sum_{n=1}^{N} (\mathbf{z}-\mathbf{\hat{z}})^{2}$. The goal is to teach the student's lower layers to mimic the behavior of the larger and well-trained teacher model in producing useful inner representations $\mathbf{\hat{z}}$, from the audio sample $\mathbf{x}$, so that, when the upper layers of the student model are trained on these high level representations, the overall performance of the wake-word system improves.

After pretraining, we add the student's upper layers (as with transfer learning) on top of the student's lower layers and train the whole system on wake word detection LF-MMI objective in Eq. \ref{eq:lfmmi_loss}. In our setup, the lower layer weights are frozen during this stage, as we found that fine-tuning the lower layers always led to degradation in performance.

\section{Experiments}
\label{sec:exp}

\begin{table*}[htb]
    \caption{False negative rate (FNR\%) at false positives per hour = 0.1 for various numbers of positive training examples, where Phone-align = phone-aligned training targets, T/S = Teacher/Student. The lowest error rate in a column is shown in bold. X indicates no discrimination of pos/neg utterances. For each model, the number of parameters and the input context (-left+right) is given in parentheses. The latency of the models is shown for the 90th percentile (in seconds).}
    \vspace{-0.2cm}
    \centering
    \resizebox{\linewidth}{!}{
      \begin{tabular}{| l |c c c c c | c | c c c c c | c |}
        \hline
        \textbf{Method} & \multicolumn{6}{c|}{\textbf{Hey Snips}}  & \multicolumn{6}{c|}{\textbf{Hey / Okay Fluency}}\\
                        &100 &500 &1000 &2000 &5799 &Latency 90\%    &100 &500 &1000 &2000 &7169 &Latency 90\%\\
        \hline
        E2E Wang et al. \cite{wang2020} (150k, -42+42)               &69.95 &8.78 &1.94 &1.82 &0.36 &1.01+0.42 s   &- &- &- &- &- &-\\
        E2E (368k, -150+10)         &53.34 &12.14 &6.13 &4.11 &\textbf{0.28} &1.00+0.10 s  &- &- &- &- &X &-\\
        E2E + Transfer (318k, -78+10)   &46.03 &18.13 &8.30 &0.99 &0.43 &0.98+0.10 s         &- &- &- &- &- &-\\
        \hline
        Phone-align (368k, -150+10)         &30.45 &12.57 &3.88 &4.11 &0.95 &0.13+0.10 s    &99.35 &24.5 &10.18 &10.68 &9.17 &0.15+0.10 s\\
        Phone-align + Transfer (318k, -78+10)              &23.92 &10.32 &4.07 &3.99 &1.94 &0.14+0.10 s    &91.75 &19.17 &5.10 &3.96 &1.59 &0.14+0.10 s\\
        Phone-align + T/S (368k, -150+10)    &\textbf{3.28} &\textbf{1.11} &\textbf{0.28} &\textbf{0.43} &0.32 &0.15+0.10 s    &\textbf{18.83} &\textbf{5.05} &\textbf{3.23} &\textbf{3.59} &\textbf{0.78} &0.14+0.10 s\\
        \hline
      \end{tabular}
      \label{tab:fnr_at_0.1fpph}
  }
  \label{tab:errorrate}
  \vspace{-0.2cm}
\end{table*}

The decoding graphs for the following experiments are tuned to a fixed false positive rate of one false positive per 10 hours, or 0.1 false positives per hour. We then report the corresponding false negative rate, the failure to recognize a spoken wake word.

\subsection{Recognition Accuracy}
\begin{figure}[t]
    \vspace{-0.2cm}
    \centering
    \begin{subfigure}[]{0.9\linewidth}
        \includegraphics[width=\linewidth]{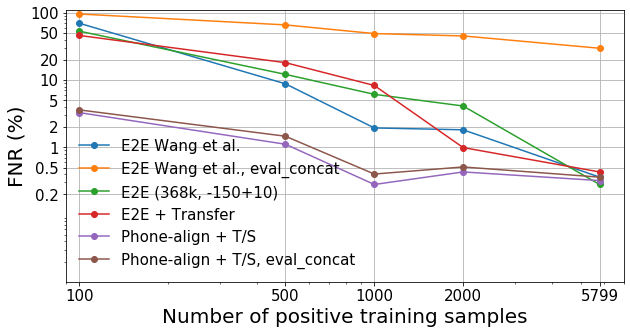}
    \end{subfigure}

    \begin{subfigure}[]{0.9\linewidth}
        \includegraphics[width=\linewidth]{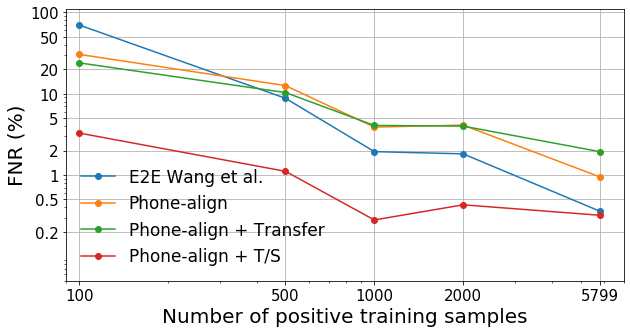}
    \end{subfigure}
    \vspace{-0.2cm}
    \caption{Snips dataset E2E (top) and Phone-align (bottom): \%FNR (log scale) vs number of training samples (log scale) at FP per hour = 0.1. Legend with ``eval\_concat" are evaluated on the eval\_concat set whereas others are evaluated on the original Snips eval set.}
    \label{fig:snips_low_resource}
    \vspace{-0.6cm}
\end{figure}

\begin{figure}[t]
\vspace{-0.2cm}
    \centering
    \resizebox{0.9\linewidth}{!}{
    \includegraphics[width=\linewidth]{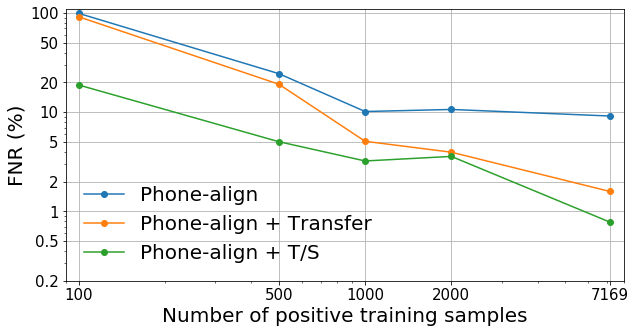}
    }
    \vspace{-0.2cm}
    \caption{Fluency dataset Phone-align: \%FNR (log scale) vs number of training samples (log scale) at FP per hour = 0.1.}
    \label{fig:fluency_low_resource}
    \vspace{-0.4cm}
\end{figure}

\vspace{-0.1cm}
Table \ref{tab:errorrate} and Figures \ref{fig:snips_low_resource} and \ref{fig:fluency_low_resource} show the performance for the various training techniques. Somewhat surprisingly, while the end-to-end-trained models performed well on the Snips dataset, end-to-end training was unsuccessful for the Fluency dataset.

To understand why, we note that the Snips dataset's positive utterances contain isolated wake words, while the Fluency dataset's positive utterances contain a wake word followed by a virtual assistant request. A second version of the Snips dataset's positive utterances were constructed to more realistically model assistant interactions: the positive utterances were replaced with ``WakeWord Speech" utterances, constructed from each positive utterance concatenated with a negative utterance from the same speaker. Performance in this "eval\_concat" condition was much worse as shown in Fig. \ref{fig:snips_low_resource}. An analysis of the output unit activations suggests that the last few HMM states of both the wake-word and speech HMMs are modeling the end of the utterance. Transitioning mid-utterance between the wake word and subsequent speech is therefore not possible. This end-of-utterance completion of the wake-word HMM also explains the high latency of the E2E models.

End-to-end training on positive utterances containing the wake word plus subsequent speech would be expected to ameliorate this problem. However, when positive utterances were formed from the concatenation of the wake word with subsequent speech, we found such training to be unsuccessful; the resultant models typically output "WakeWord Speech" for most every utterance. Similar results were observed on the Fluency dataset. We attribute this failure to the additional challenge of learning which prefix of the positive utterances constitutes the wake word. Perhaps the task would be learnable with significantly more data. 

Phone-aligned training made learning on the Fluency dataset possible. Additionally, the phone-aligned Snips model performed similarly on the ``eval-concat" version of the test set. For a fair comparison of E2E and phone-align models in terms of model capacity, besides the baseline model \cite{wang2020} which has 150k parameters and input context of -42+42 frames, we also experiment with a model the same size and architecture as that of our best performing student model, however we do not find any improvement. This model, with 368k parameters and input context of -150+10 frames, is denoted by ``E2E (368k, -150+10)".


Unsurprisingly, all approaches benefit from more training data. Transfer learning was beneficial for phone-aligned models trained on less data and continued to show benefits on the more challenging Fluency dataset, but the teacher-student pretraining consistently performed the best across both datasets.

Symmetric and asymmetric teacher-student training pipelines were compared. In the symmetric case, the same augmented audio is provided to teacher and student. In the asymmetric case, clean audio is used to generate teacher output representations, while the student model representation is generated from augmented audio. The aim here is to teach the student's lower layers not only to mimic the teacher's phonetic inference but also to learn to produce noise-robust representations. We found consistent benefits from the asymmetric approach, so the ``Phone-align + T/S" results use this approach.

Finally, we observe that fine-tuning the pretrained layers did not help. One reason may be that fine-tuning on general speech (whose 4 or 5 states in the ``Speech" HMM do not have good alignment with the same phones) may lead to training signals that cause the pretrained layers to unlearn its phone recognition capabilities. In future, we will explore using multi-task learning -- teacher-student loss + wake-word recognition -- similar to the transfer learning approach of \cite{sun2017,gao2020}.

\subsection{Latency Analysis}
\vspace{-0.1cm}
For phone-aligned training, the final state of the HMM is trained to align to the first 10 frames \emph{after} the wake word, which is usually silence. This encourages the model to only wait to see 10 frames (or 100ms) following the wake word in order to trigger. Table \ref{tab:fnr_at_0.1fpph} shows the 90th percentile latency of models using end-to-end and phone-aligned training for the largest dataset size. ASR alignments were used as the reference for the precise end of each wake word. Consistent with the training targets, phone-aligned models show a latency of only 130-150ms, though the additional latency due to future feature frames relative to the model’s target prediction frame (given after ``+'' sign) should be considered as part of the total user-experienced latency. Even without considering the frame look-ahead, E2E models' latency is still much higher.

\vspace{-0.2cm}
\section{Conclusions}

We compared the performance of a strong baseline system trained on various amounts of wake-word data to models trained with either of two pretraining techniques leveraging speech recognition acoustic models: transfer learning and knowledge distillation. Compared to the baseline system and a transfer learning model, knowledge distillation performed better across both datasets and across dataset sizes, with a particularly dramatic error rate reduction when wake-word data was more limited. 

Additionally, we found that phone-aligned training was able to reduce  latency to less than 250ms, and is necessary to train a wake-word model on the more challenging Fluency dataset.

In future, we will explore these techniques on a wider range of model architectures, including simplified output representations, as well as further explore strategies for fine-tuning the lower pre-trained layers of the network.


\bibliographystyle{IEEEtran}
\bibliography{main}


\end{document}